# Giant Rashba-Splitting of One-Dimensional Metallic States in Bi Dimer Lines on InAs(100)

Polina M. Sheverdyaeva[1], Gustav Bihlmayer[2], Silvio Modesti[3,4], Vitaliy Feyer[5], Matteo Jugovac[1,5], Giovanni Zamborlini[5,6,7], Christian Tusche[5,8], Ying-Jiun Chen[5,8], Xin Liang Tan[5], Kenta Hagiwara[5], Luca Petaccia[9], Sangeeta Thakur[9,10], Asish K. Kundu[1,11,12], Carlo Carbone[1], and Paolo Moras[1]

[1]Istituto di Struttura della Materia-CNR (ISM-CNR), Strada Statale 14 km 163.5, 34149 Trieste, Italy
[2]Peter Grünberg Institut (PGI-1), Forschungszentrum Jülich and JARA, D-52425 Jülich, Germany
[3]Dipartimento di Fisica, Università di Trieste, 34127 Trieste, Italy
[4]Istituto Officina dei Materiali, Consiglio Nazionale delle Ricerche, 34149 Trieste, Italy
[5]Peter Grünberg Institute (PGI-6), Forschungszentrum Jülich, 52428 Jülich, Germany
[6]TU Dortmund University, Otto-Hahn-Straße 4, 44227 Dortmund, Germany
[7]Institute of Physics, University of Graz, Universitätsplatz 5, 8010 Graz, Austria
[8]Fakultät für Physik, Universität Duisburg-Essen, 47057 Duisburg, Germany
[9]Elettra Sincrotrone Trieste, Strada Statale 14 km 163.5, 34149 Trieste, Italy
[10]Freie Universität Berlin, Institut für Experimentalphysik Arnimallee 14, 14195 Berlin, Germany
[11]Condensed Matter Physics and Materials Science Department, Brookhaven National Laboratory, Upton, New York 11973, USA
[12]International Center for Theoretical Physics (ICTP), 34151 Trieste, Italy

**Abstract**

Bismuth produces different types of ordered superstructures on the InAs(100) surface, depending on the growth procedure and coverage. The (2×1) phase forms at completion of a Bi monolayer and consists of a uniformly oriented array of parallel lines of Bi dimers. Scanning tunneling and core level spectroscopies demonstrate its metallic character, in contrast with the semiconducting properties expected on the basis of the electron counting principle. The weak electronic coupling among neighboring lines gives rise to quasi one-dimensional Bi-derived bands with open contours at the Fermi level. Spin- and angle-resolved photoelectron spectroscopy reveals a giant Rashba splitting of these bands, in good agreement with ab-initio electronic structure calculations. The very high density of the dimer lines, the metallic and quasi one-dimensional band dispersion and the Rashba-like spin texture make the Bi/InAs(100)-(2×1) phase an intriguing system, where novel transport regimes can be studied.

**Introduction**

The Rashba-Bychkov (RB) effect lifts the spin degeneracy of the electronic bands in crystalline solids with broken structural inversion symmetry [1], typically caused by the presence of a surface or an interface. This effect has attracted great attention as a fundamental mechanism for spin generation/control in spintronic devices [2,3], owing to the spin-momentum locking, which constraints the spin and momentum directions of an electron to be mutually perpendicular. The RB effect has been detected by angle-resolved photoelectron spectroscopy (ARPES) in many two-dimensional (2D) systems and classified according to the Rashba parameter $\alpha_R$, which quantifies the energy-momentum separation of the bands with opposite spins [4-7]. $\alpha_R$ depends on the atomic spin-orbit coupling (SOC) and the gradient of the electron potential across the structural discontinuity (surface/interface plane) [8-11]. The giant RB effect ($\alpha_R > 3$ eV·Å) has been suggested to play a key role in spin-to-charge conversion phenomena occurring in 2D heterostructures [12-15]. The discovery of a large Rashba splitting in the metallic states of Pb on Ge(111) has opened the way to spin accumulation, filtering and injection in semiconductor materials [16].

The RB effect can significantly influence the electronic structure and spin texture of one-dimensional (1D) systems, such as quantum wires, and favor the emergence of a specific spintronic functionality [17]. In a 1D system subject to the RB interaction the spin degeneracy of the spin-split bands at the time-reversal symmetry point can be removed by an external magnetic field opening a gap. If this gap opens at the Fermi level ($E_F$), a pure and non-dissipative spin current can be established by applying a voltage. The exploitation of this mechanism is not trivial due to the scarcity of systems with genuine 1D RB-like spin texture. As an example, the 1D bands of Bi chains on Ag(110) display giant RB splitting, but the density of states of the system at $E_F$ is dominated by bulk Ag states [18]. Semiconductor substrates, on the other hand, can support the formation 1D structures with RB-split metallic states, as experimentally demonstrated for Au chains on Si(557) [19], and theoretically predicted for Bi-adsorbed In atomic chains on Si(111) [20].

Bi is known to form different types of superstructures on III-V semiconducting surfaces [21-34]. Some of these are based on Bi stripes displaying quasi 1D bands with giant RB splittings [28,30-32,34]. Bi-terminated III-V semiconductors are expected to be non-metallic, according to the electron counting model, in order to decrease the surface energy [35]. A notable exception to this model is the metallic behavior of Bi dimer lines grown on GaAs(100) (the so-called (2×1) phase), which emerges from scanning tunneling spectroscopy (STS) measurements and finds confirmation in density functional theory (DFT) calculations (without SOC) [23, 24]. An in-depth analysis aimed at establishing the dimensionality of the Bi-derived states and the magnitude of the RB effect in this and similar systems (Bi/GaAs$_x$N$_{1-x}$(100)-(2×1) [22] and Bi/InAs(100)-(2×1) [26]) is still missing.

The present study reports on the electronic structure of the Bi/InAs(100)-(2×1) phase, which is examined by STS, photoelectron spectroscopy with spin analysis and DFT calculations. The (2×1) phase occurs at the completion of one Bi monolayer and consists of an array of uniformly oriented Bi dimer lines. The electronic coupling between neighboring lines turns out to be much weaker than along the dimer lines, thus giving rise to Bi-derived bands with highly anisotropic in-plane dispersion. The Fermi surface presents open quasi 1D contours with RB-like spin texture and giant splitting with α$_R$ values up to 4.6 eV·Å, owing to the large SOC of Bi and the low structural symmetry. These findings suggest that the Bi/InAs(100)-(2×1) phase could support the generation of non-dissipative and spin-polarized currents and find application in spin-to-charge conversion processes.

**Methods**

Undoped n-type wafer substrates of InAs(100) were treated with Ar ion sputtering (800 eV) and annealing (720 K) cycles. This procedure results in the formation of the (4×2)/c(8×2) surface reconstruction visible in the low-energy electron diffraction (LEED) pattern of Fig. 1(a) [26,36,37]. Bi was deposited from a resistively heated source in excess to one monolayer (1 ML corresponds to one Bi atom per (1×1) surface unit cell of InAs(100)) on the substrate kept at 300 K. The Bi/InAs(100) interface displays the following sequence of ordered superstructures for decreasing Bi coverage: (2×6) for more than 1.33 ML Bi [21,29,33], (2×10) at about 1.2 ML Bi [26], (2×1) at 1 ML Bi [26], (2×8) coexisting with (2×4) below 1 ML Bi [24], (2×4) at 0.38 ML Bi [21]. 60-minute-long annealing at 550 K was necessary to release Bi exceeding 1 ML from the surface and stabilize the (2×1) phase (LEED pattern in Fig. 1(b)). This phase consists of a uniform array of Bi dimer lines running along the [011] direction, as shown in the scanning tunneling microscopy (STM) image of Fig. 1(c)) [26]. STM and STS data were acquired with a home-made instrument at 300 K for the clean InAs(100) surface and at 70 K for the Bi/InAs(100)-(2×1) phase by using a gold tip. The bias voltage and current were set to of V$_{bias}$ = -0.5 V and I = 3×10$^{-11}$ A to acquire constant current topographic maps. The differential conductance (dI/dV) curves were measured by the lock-in technique with a 20 mV modulation of V$_{bias}$. Photoelectron spectroscopy measurements were carried out at the VUV-Photoemission, BaDElPh [38] and NanoESCA [39] beamlines of the Elettra Synchrotron (Trieste, Italy) at liquid nitrogen temperature. Core level and

ARPES spectra were collected with hemispherical electron spectrometers at the VUV-Photoemission and BaDElPh beamlines. Spin- and momentum-resolved constant energy maps of the photoelectron signal were acquired at the NanoESCA beamline using a momentum microscope, which is equipped with a W(100)-based spin detector [40]. The analysis of the spin-resolved maps was performed following the procedure described in Ref. [41].

DFT calculations were performed in the local density approximation [42] using the full potential linearized augmented plane-wave method [43]. The Bi/InAs(100)-(2×1) system was simulated with a slab of 21 atomic layers based on the symmetric dimer model proposed for the Bi/GaAs(100)-(2×1) system [24]: one-atom-thick Bi dimer lines running along the [011] direction on the top surface; 19 atomic layers of InAs(001), with outermost In planes, as the substrate; hydrogen-terminated bottom surface. Fig. 1(d,e) display the top view and the three-dimensional representation of this structural model. The in-plane lattice constant (4.27 Å) and interlayer distance between In (or As) planes (3.02 Å) were set to those of bulk InAs. Structural relaxation on the Bi-terminated side of the slab led to 2.96 Å Bi-In interlayer distance, 3.07 Å Bi-Bi distance in the dimers and 0.5% expansion of the interlayer distance between the topmost In planes. Hydrogen atoms were placed on the bottom surface to saturate the dangling bonds. An attractive potential term of 2.67 eV was applied to the *p*-states of In and As to get a better description of the band gap and SOC was included in a self-consistent manner. The In terminations on both sides of the InAs substrate led to an artificial hole doping of the system and to a difference of 0.22 eV between the $E_F$ position in the calculations and experiments.

**Results and Discussion**

Fig. 1(c) shows a typical STM image acquired on a sample displaying the (2×1) LEED pattern (Fig. 1(b)). The observation of light gray rectangles with 2:1 ratio between the long side (aligned with the [0$\bar{1}$1] substrate axis) and the short side (aligned with the [011] substrate axis) is a signature of the dimerization of Bi atoms, which are not resolved individually in the image [26]. The dimers form an array of parallel lines running along the [011] axis. The continuity of the Bi dimer lines is occasionally interrupted by missing dimers (dark gray areas). The structural features observed in the STM image are reproduced by the model of the Bi/InAs(100)-(2×1) phase. Bi atoms forming dimers present a reduced Bi-Bi distance along [0$\bar{1}$1] with respect to the surface lattice constant of the substrate (Fig. 1(d)). These Bi dimers give rise to the parallel lines seen in Fig. 1(e).

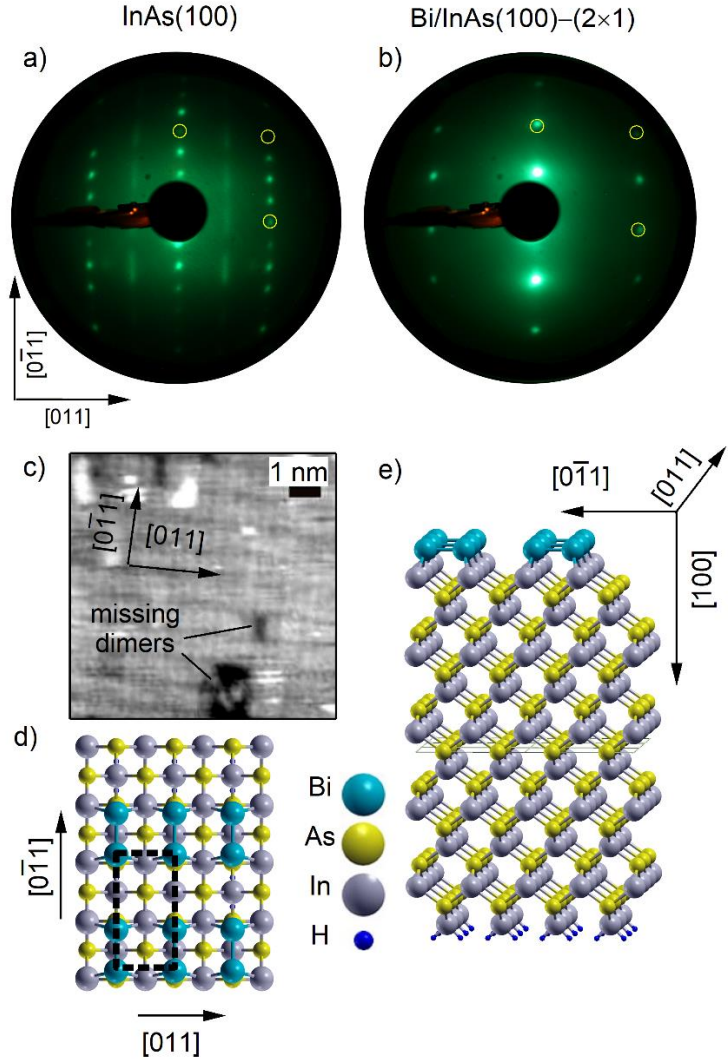

**Fig. 1** *(a) LEED pattern of clean InAs(100) with (4×2)/c(8×2) reconstruction at 48 eV primary electron energy. Yellow circles mark the position of the (1×1) spots. (b) LEED pattern of the Bi/InAs(100)-(2×1) phase at 46 eV primary electron energy. Yellow circles mark the position of the (1×1) spots. (c) Constant current STM image of the Bi/InAs(100)-(2×1) phase. Light gray rectangles represent the Bi dimers. Missing dimers appear as dark regions. (d) Top view and (e) three-dimensional rendition of the structural model used to calculate the electronic structure of the Bi/InAs(100)(2×1) phase. The dashed rectangle in (d) represents the (2×1) surface unit cell.*

Fig. 2 compares STS and photoelectron spectra of the clean InAs(100) surface and Bi/InAs(100)-(2×1) phase. The dI/dV signal of clean InAs(100) (top spectrum of Fig. 2(a)) is close to zero over an interval of 0.30 eV including $E_F$ that can be identified with the bulk bandgap of the substrate [44]. The dI/dV signal of the Bi/InAs(100)-(2×1) phase (collected away from missing dimer regions, bottom spectrum of Fig. 2(a)) is more than a factor 10 higher than the noise level at $E_F$, thus attesting the metallic character of the (2×1) phase. The peaks observed at -0.2, -0.08 and 0.08 eV (black arrows) will be discussed later in connection with the ARPES and DFT analyses.

Survey spectra of the photoelectron signal at hν = 70 eV for the two systems (Fig. 2(b)) are useful to compare the intensities of the core level lines. In both cases, the much higher signal of In with respect to As has combined structural and electronic origin, as the sputtering/annealing procedure results in an excess of In at the surface and In *4d* states have 5.6 times larger photoionization cross section than As *3d* states [45]. The zoom of the photoelectron signal near $E_F$ for InAs(100) (top spectrum) and Bi/InAs(100)-(2×1) (bottom spectrum) in Fig. 2(c) highlights the semiconducting vs metallic properties of the two systems.

Fig. 2(d-f) show the analysis of the As *3d*, Bi *5d* and In *4d* levels of InAs(100) (top spectra) and Bi/InAs(100)-(2×1) (bottom spectra), respectively. The As *3d* spectrum of clean InAs(100) can be fitted with two doublets related to sub-surface As atoms (red line, As *3d$_{5/2}$* at 40.50 eV) and bulk-like As atoms (green line, As *3d$_{5/2}$* at 40.69 eV). The sub-surface component is suppressed due to a restructuring of the (4×2)/c(8×2) termination, occurring in correspondence with the formation of the (2×1) phase. The Bi *5d* spectrum can be fitted with one doublet (blue line, Bi *5d$_{5/2}$* at 23.55 eV) very similar to that of metallic Bi [46], at variance with the two doublets used in the literature [26]. The slight asymmetry of the peaks (tail on the high binding energy side) is interpreted as a signature of the metallic character of the Bi/InAs(100)-(2×1) phase [47], rather than ascribed to the presence of another phase [26]. This interpretation is in agreement with the data of Fig. 2(a,c) and will be strengthened by the ARPES and DFT analyses reported in Fig. 3 and 4. Two doublets corresponding to surface (red line, In *4d$_{5/2}$* at 17.48 eV) and bulk (green line, In *4d$_{5/2}$* at 17.23 eV) components are sufficient to fit the In *4d* spectrum of clean InAs(100). After the formation of the (2×1) phase, the surface doublet is quenched and another doublet appears on the low binding energy side of the bulk component. This new doublet has an asymmetric shape (blue line, In *4d$_{5/2}$* at 16.93 eV) that is compatible with metallic In [48] in contact with surface metallic Bi.

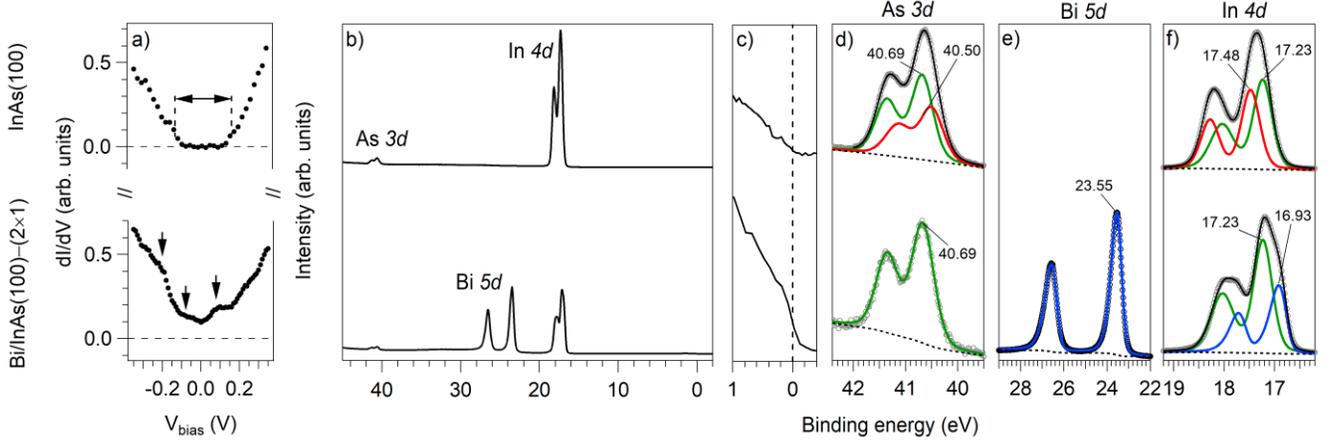

**Fig. 2** *(a) dI/dV spectra for clean InAs(100) at 300 K (top spectrum) and Bi/InAs(100)-(2×1) at 70 K (bottom spectrum). Negative $V_{bias}$ values indicate occupied states. (b) Survey photoelectron spectra at hv = 70 eV of clean InAs(100) (top spectrum) and Bi/InAs(100)-(2×1) (bottom spectrum) and (c) corresponding spectra near $E_F$. (d-f) Core level spectra hv = 70 eV and related fittings of the (e) As 3d (f) Bi 5d and (g) In 4d lines for the clean InAs/(100) surface (top) and Bi/InAs(100)-(2×1) phase (bottom).*

Fig. 3(a-c) show a constant energy cut of the photoelectron signal of the Bi/InAs(100)-(2×1) phase acquired at 0.15 eV below $E_F$ with the momentum microscope. This instrument allows to scan simultaneously an area of the ($k_x$,$k_y$) space including several surface Brillouin zones (SBZs) of the system. For clarity, the edges of the (2×1) SBZs (black dashed lines) are overlaid to the data. The central SBZ (rectangle with thick black edges, $\overline{\Gamma X}$ = 0.74 Å$^{-1}$ and $\overline{\Gamma Y}$ = 0.37 Å$^{-1}$) and its high symmetry points are reported in Fig. 3(b). The photoelectron signal is characterized by four bands $S_1$-$S_4$, which are elongated in the $k_y$ direction and cross the edges of neighboring SBZs. Fig. 3(c) displays the second derivative of the original data taken along the $k_x$ axis, to better visualize low intensity sections of $S_1$-$S_4$. All the features observed in Fig. 3(a-c) are absent in the clean substrate (Fig. S1 of the Supplemental Information file) [49], reflect the periodicity of the (2×1) superstructure, and, therefore, can be identified as Bi-derived electronic states. The wavy $S_1$/$S_2$ and straight $S_3$/$S_4$ constant energy contours are schematically represented by red and blue continuous lines for positive $k_x$ values in Fig. 3(b), where the color is assigned on the basis of the spin analysis.

In the spin-resolved constant energy map of Fig. 3(d) the spin quantization axis (SQA), determined by the experimental geometry [39], is parallel to the $k_y$ axis. The intensity of red and blue colors for spin-up and spin-down states, respectively, is proportional to their spin projection along the SQA. The most evident feature of Fig. 3(d) is the full reversal of the spin texture with respect to the $k_y$ axis. Portions of the $S_1$-$S_3$ contours are highlighted by red/blue dashed lines for positive $k_x$ values to guide the eye. $S_1$ and $S_2$ display high and opposite spin polarizations, thus suggesting they form a RB pair. The spin analysis of $S_3$ and $S_4$ is hindered by their low intensity close to $k_x = 0$ axis, while their high spin-polarization clearly emerges at larger $|k_x|$ values. The high spin-polarization of the $S_1$-$S_4$ states in Fig. 3(d) means that the spins of the Bi-derived bands are almost parallel or anti-parallel to the SQA.

The ARPES spectra of Fig. 3(e-i) show the energy-momentum dispersion of the $S_1$-$S_4$ bands along the segments marked by the green dashed lines in Fig. 3(b). Also in this case the data are presented in the second derivative form to enhance the sensitivity to weak features. The three ARPES maps taken along equivalent $\bar{X}$–$\bar{\Gamma}$–$\bar{X}$ directions (Fig. 3(e-g)) demonstrate the strong state-dependent modulation of the photoelectron signal in the ($k_x$,$k_y$) plane, due to matrix element effects. $S_1$ and $S_2$ clearly cross $E_F$, thus making the Bi/InAs(100)-(2×1) phase metallic, in agreement with the analysis displayed in Fig. 2. This metallic character is a distinctive feature of the (2×1) phases that 1 ML Bi forms on GaAs(100) [23,24] and GaAs$_x$N$_{1-x}$(100) [22]. All $S_1$-$S_4$ states can be observed simultaneously along the $\bar{S}$–$\bar{Y}$–$\bar{S}$ direction of Fig. 3(h). The dispersion of $S_3$ and $S_4$ (Fig. 3(g,h) and Fig. S2 of the Supplemental Information file) closely reminds the RB-split states observed in the (2×1) phases of Bi on the (110) surfaces of III-V semiconductors [28,30-32]. This suggests that also $S_3$/$S_4$ form a RB pair of bands, in analogy to $S_1$/$S_2$. Fig. 3(i) shows the ARPES spectra along the $\bar{Y}$–$\bar{\Gamma}$–$\bar{Y}$ direction. The flat feature represents the $S_3$/$S_4$ crossing point. The minimum at $\bar{\Gamma}$ (0.2 eV) and the flat dispersion about $\bar{Y}$ (0.08-0.1 eV) can be directly correlated to the peaks observed at the same energies in the occupied part of the STS spectrum in Fig. 2(a). The STS peak at 0.08 eV above $E_F$ can be linked to the maxima of the $S_3$ and $S_4$ bands (see Fig. 4), which cross $E_F$ and, therefore, contribute to the metallicity of the system.

The data of Fig. 3 clearly demonstrate that all $S_1$-$S_4$ bands have an anisotropic in-plane dispersion, which can be associated to the structural properties of the (2×1) phase. Due to the formation of the Bi dimers, the separation between Bi atoms in neighboring dimer lines is larger than along the lines. Correspondingly, the electronic coupling of Bi-related states along the dimer lines is stronger that perpendicular to them, thus resulting in steeply dispersing and flat bands, respectively, and open elongated contours. Overall, the ARPES and with the spin polarization analysis allow to describe $S_1$/$S_2$ and $S_3$/$S_4$ as RB-split pairs of bands with quasi 1D and metallic character.

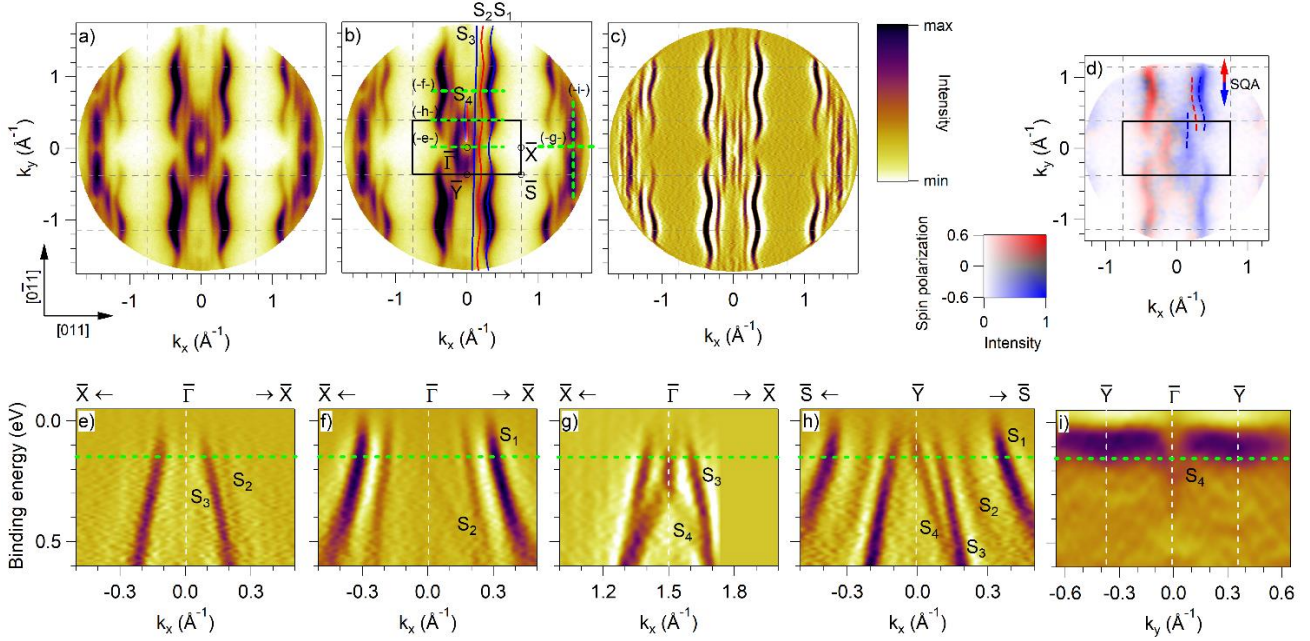

**Fig. 3**. *(a-c) Constant energy cuts of the photoelectron signal for the Bi/InAs(100)-(2×1) phase taken at 0.15 eV binding energy with hv = 65 eV and p-polarized light. Black dashed lines indicate the edges of the SBZs. (a) Original data. (b) High-symmetry points of the central SBZ (thick black line), and constant energy contours of the $S_1$-$S_4$ bands (red/blue continuous lines) are plotted on the original data. (c) Second derivative of the original data along the $k_x$ axis. (d) Spin-polarized constant energy cut at 0.15 eV. SQA and SBZs (dashed lines) are shown. (e-i) ARPES spectra collected along the green dashed lines shown in panel (b). All data are displayed in the second derivative form. (e-g) Spectra along three equivalent $\bar{X}$–$\bar{\Gamma}$–$\bar{X}$ directions. (h) Spectra along the $\bar{S}$–$\bar{Y}$–$\bar{S}$ direction. (i) Spectra along the $\bar{Y}$–$\bar{\Gamma}$–$\bar{Y}$ direction.*

In order to interpret the experimental findings described above, the electronic structure of the Bi/InAs(100)-(2×1) system was computed by DFT using the model reported in Fig. 1(d,e), which is based on the symmetric dimer model proposed for the Bi/GaAs(100)-(2×1) system [24]. Fig. 4(a) shows the spin-resolved DFT band structure calculations along the high-symmetry directions of Bi/InAs(100)(2×1). The size of the symbols is proportional to the spin polarization of the states, with the SQA oriented along the $k_y$ axis. Red/blue colors correspond to up/down spin channels, in analogy to the experiment. The horizontal dashed line indicates the experimental position of $E_F$, which lies 0.22 eV above $E_F$ in the calculations. This difference has been evaluated by aligning the experimental and calculated $S_3$/$S_4$ bands (Fig. 4(b)). It derives from the artificial hole doping due to the excess of In atoms in the model (10 In planes vs 9 As planes). To ease the comparison between experiment and theory, the energy scale of the calculations will be referred to the experimental position of $E_F$ (right axis of Fig. 4(a)) from here onwards. The electronic structure in the proximity of $E_F$ is characterized by four spin-polarized and Bi-derived bands, which are labeled with $S_1$-$S_4$ in analogy to the experiment. The additional bands crossing $E_F$ along $\bar{\Gamma}$–$\bar{X}$ and $\bar{\Gamma}$–$\bar{Y}$ and showing no spin-polarization originate from the H-terminated surface and are not relevant to the present discussion.

All $S_1$-$S_4$ bands display quasi 1D character, which emerges by comparing their steep dispersion along $\bar{\Gamma}$–$\bar{X}$ and $\bar{Y}$–$\bar{S}$ (parallel to the Bi dimer lines) with their flat dispersion along $\bar{\Gamma}$–$\bar{Y}$ and $\bar{X}$–$\bar{S}$ (perpendicular to the Bi dimer lines). The spin analysis reveals that $S_1/S_2$ and $S_3/S_4$ form two RB pairs. The strength of the RB effect can be evaluated in the proximity of $\bar{\Gamma}$ and $\bar{Y}$ through $\alpha_R = 2\cdot\Delta E/\Delta k$, as shown in Fig. 4(b). The $S_1/S_2$ pair presents giant $\alpha_R$ values of 4.6 eV·Å along $\bar{S}$–$\bar{Y}$–$\bar{S}$ and 3.5 eV·Å along $\bar{X}$–$\bar{\Gamma}$–$\bar{X}$. For the $S_3/S_4$ pair $\alpha_R$ is 2.3 eV·Å along $\bar{S}$–$\bar{Y}$–$\bar{S}$ and 2.7 eV·Å along $\bar{X}$–$\bar{\Gamma}$–$\bar{X}$. In Fig. 4(b) the calculated bands are overlaid to the ARPES data of Fig. 3(h) to demonstrate the correspondence between theory and experiment. The slight offset of the experimental $S_1/S_2$ pair with respect to the calculations can indicate that the spin-splitting and, consequently, the $\alpha_R$ value are larger than predicted. A similar effect is seen Fig. S2 of the Supplemental Information file for the $S_3/S_4$ pair. The overall good agreement allows to describe the Bi/InAs(100)-(2×1) phase as an array of Bi dimer lines with metallic character determined by two pairs of bands displaying quasi 1D dispersion and RB-like spin texture.

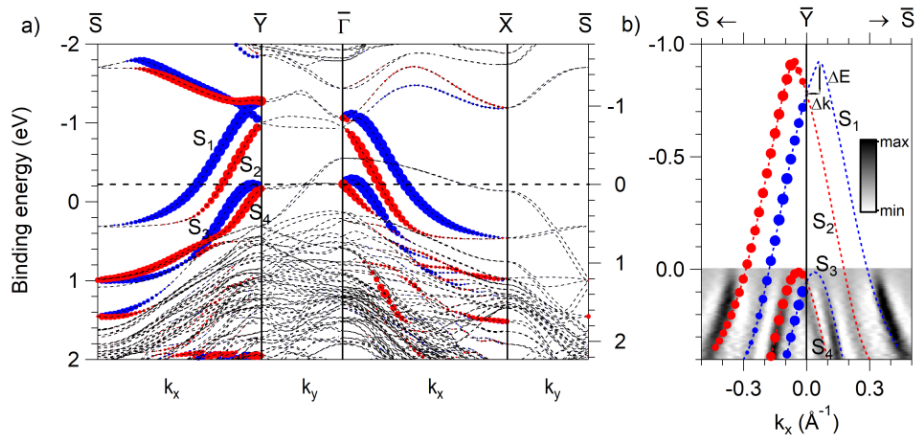

**Fig. 4**. *(a) Spin-resolved DFT calculations. The size of the symbols is proportional to the in-plane component of the spin polarization along the $k_y$ axis. The location of $E_F$ in the calculation (left axis) and in the experiment (right axis) differ by 0.22 eV. (b) Comparison between the band dispersion calculated by DFT and measured by ARPES along the $\bar{S}$–$\bar{Y}$–$\bar{S}$ direction using the experimental energy scale.*

The analysis reported above suggests different types of spintronic functionalities related to the specific properties of the $S_1/S_2$ and $S_3/S_4$ RB pairs. The robust metallic character and giant RB parameter of the $S_1/S_2$ pair can find application in spin-to-charge conversion, whose efficiency is expressed by $\lambda = \alpha_R \tau_s / \hbar$, where $\tau_s$ is the spin-relaxation time [50]. Notably, the decoupling of the $S_1/S_2$ surface bands from the bulk states of the semiconductor substrate near $E_F$ [15] and the reduced back-scattering associated to the quasi 1D dispersion [51-53] favor large $\tau_s$ values. From the fundamental point of view, the $S_1/S_2$ bands could be also used for studying exotic electronic phenomena, such as Majorana bound states [54] and spin-dependent density waves [55]. At variance with the $S_1/S_2$ bands, the maxima of the $S_3/S_4$ pair lie just above $E_F$ close to $\bar{\Gamma}$. Most likely these maxima produce the peak located at 0.08 eV in the unoccupied part of the STS spectrum (Fig. 2(a)). In fact, $S_1$ and $S_2$ are expected to give rise to a featureless dI/dV signal, due to their steep dispersion and distance from $\bar{\Gamma}$ near $E_F$. The vicinity of the $S_3/S_4$ crossing point to $E_F$ at $\bar{\Gamma}$, which is a time-reversal symmetry point of the system, can realize the scenario depicted in Ref. [17]: An external magnetic field could open a gap between $S_3$ and $S_4$; if the gap includes $E_F$, the application of a voltage would allow the flow of a non-dissipative, pure spin current through the Bi dimer lines. Notably, the exact location of the $S_3/S_4$ bands with respect to $E_F$ can be tuned by external doping or gating, thanks to the semiconducting nature of the substrate, to meet the requirements of Ref [17].

**Conclusion**

The present work reports on the electronic structure of the Bi/InAs(100)-(2×1) phase. This system can be described as a compact array of Bi dimer lines, whose metallic character is determined by two pairs of quasi 1D bands displaying RB-type spin texture. The location of the low-lying $S_3/S_4$ pair, with a crossing point at $\bar{\Gamma}$ very close to $E_F$, appears to be suitable to host pure spin-polarized and non-dissipative currents, upon the application of an external magnetic field. Additional spin-to-charge conversion functionalities could be introduced by the $S_1/S_2$ pair with giant RB splittings. The experimental verification of these novel transport properties would open new perspectives for the exploitation of the RB effect in spintronic devices.

**Data availability**

Data will be made available on request.

**Author contributions**

P. M. Sheverdyaeva: investigation, data curation, validation, formal analysis, writing – original draft, project administration. G. Bihlmayer: resources, software, conceptualization, validation, writing – review and editing. S. Modesti: investigation, data curation, validation, formal analysis, writing – review and editing. V. Feyer, M. Jugovac, G. Zamborlini, C. Tusche, Ying-Jiun Chen, X. L. Tan, K. Hagiwara, L. Petaccia, S. Thakur, A. K. Kundu: investigation, data curation, validation, writing – review & editing. Carlo Carbone: validation, writing – review & editing. Paolo Moras: conceptualization, supervision, validation, writing – review & editing. All authors participated in the analysis and discussion of the results.

**Conflicts of interest**

There are no conflicts to declare.

**Acknowledgments**

We acknowledge EUROFEL-ROADMAP ESFRI of the Italian Ministry of University and Research. G.B. gratefully acknowledges the computing time granted through JARA-HPC on the supercomputer JURECA at Forschungszentrum Jülich. A.K.K. acknowledges receipt of a fellowship from the ICTP-TRIL Programme, Trieste, Italy. C.T. and Y.-J.C. acknowledge support by the German Federal Ministry of Education and Research (BMBF) under grant No. 05K19PGA. We acknowledge Elettra Sincrotrone Trieste for providing access to its synchrotron radiation facilities (beamtime nr. 20170258, 20175193, 20185310).

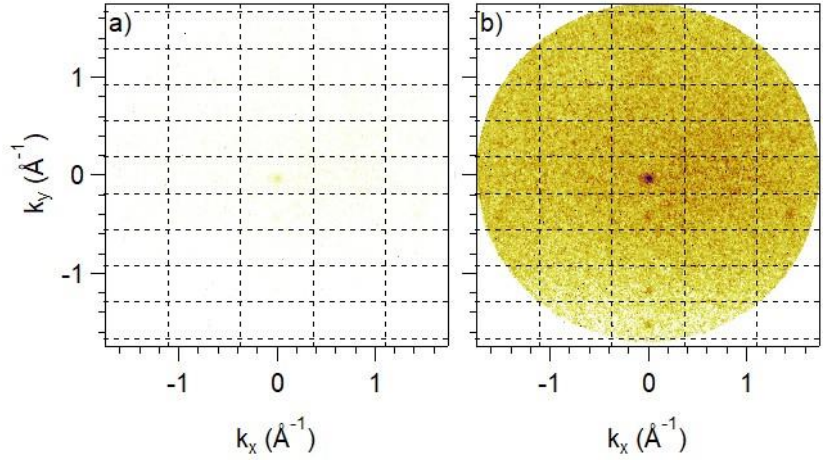

**Fig. S1**. *Constant energy cut (0.15 eV below $E_F$) of the photoelectron signal of the (4×2)/c(8×2)-terminated InAs(100) substrate taken with the momentum microscope under the same experimental conditions of Fig. 3(a). The color scale in panel (a) is the same of Fig. 3(a), while panel (b) uses a color scale that enhances very low intensity electronic features. The dashed lines indicate the edges of the (4×2) SBZ. The small circles are charge accumulation states derived from the InAs conduction band. These states are visible at some $\bar{\Gamma}$ points of (4×2) SBZs, which are also $\bar{\Gamma}$ points of the hexagonal-like c(8×2) SBZs (not shown).*

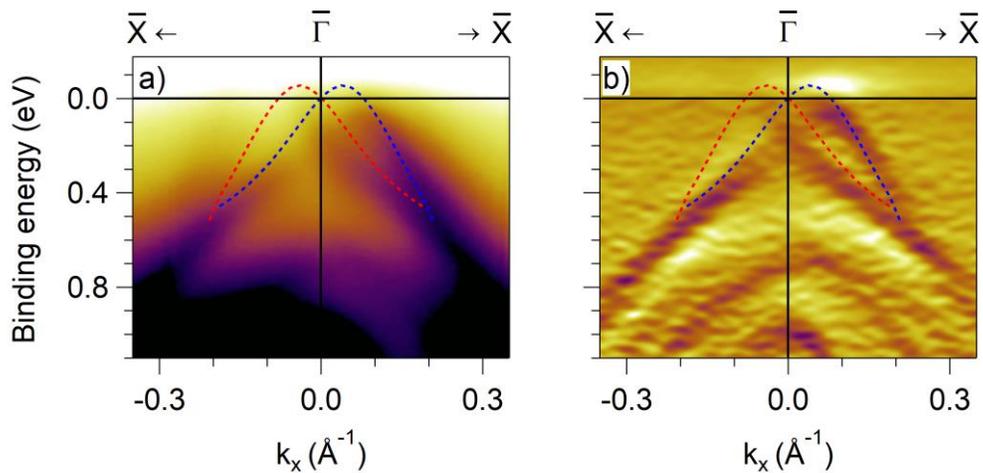

**Fig. S2**. *ARPES spectra taken at the BaDElPh beamline with hv = 22 eV along the same segment probed in Fig. 3(e) and corresponding to the $\overline{X}$-$\overline{\Gamma}$-$\overline{X}$ direction. Panels (a) and (b) report the original and second derivative data, respectively. The slight discrepancy with the calculated $S_3$ and $S_4$ bands (dashed lines) can indicate a larger $\alpha_R$ value in the experiment than in the theory.*